\documentclass[a4paper,11pt]{article}
\usepackage{jheppub} % for details on the use of the package, please
\usepackage{lmodern, physics, varioref, cleveref}

\newcommand{\dsum}{\displaystyle\sum}

\newcommand{\dint}{\displaystyle\int}
\renewcommand{\d}{\partial}
\renewcommand{\dfrac}{\displaystyle\frac}

\newcommand{\overbar}[1]{\mkern 2mu\overline{\mkern-2mu#1\mkern-2mu}\mkern 2mu}

\newcommand{\R}{\mathbb{R}}

\newcommand{\Z}{\mathbb{Z}}

\newcommand\mperiod[1][\rlap]{#1{\ .}}	%punctuation
\newcommand\mcomma[1][\rlap]{#1{\ ,}}

\crefname{table}{table}{tables}
\Crefname{table}{Table}{Tables}
\crefname{figure}{figure}{figures}
\Crefname{figure}{Figure}{Figures}

\newenvironment{eqaed}
    {\begin{equation}
    \begin{aligned}
    }
    { 
    \end{aligned}
    \end{equation}
    \ignorespacesafterend
    }
    
\title{\boldmath On Codimension-one Vacua and String Theory}
\author{Salvatore Raucci}
\affiliation{Scuola Normale Superiore and INFN,\\Piazza dei Cavalieri 7, 56126 Pisa, Italy}
\emailAdd{salvatore.raucci@sns.it}

\abstract{We investigate codimension-one vacua arising from low energy effective actions inspired by string theory, with an eye to their consistency when localized sources are allowed in the equations of motion. We draw some inspiration from Sugimoto's USp(32) model, the simplest setting for brane supersymmetry breaking, and from the 0'B model, with their Dudas-Mourad solutions. Although the sources that one can thus identify do not have a clear role in string theory, this type of investigation is naturally suggested by the singularities that appear at the endpoints of internal intervals. We also discuss the introduction of sources in deformed D8-like solutions in type IIA, pointing out an analogy with one of the non-supersymmetric models. Finally, we show that an appropriate choice of frame can simplify computations in models with tadpole potentials. }

\begin{document} 

\maketitle
\flushbottom
\newpage

\section{Introduction}\label{sec:introduction}

Superstring theory can recover results from both general relativity and quantum field theory in a consistent way, but appears to require spacetime supersymmetry in any construction that is reasonably under control.
The problem is of wide importance, since there is usually no clear separation between the consequences of supersymmetry and those of UV consistency. In order to settle these puzzling issues, one would probably need a deeper reformulation and further understanding of the string landscape. At present, the Swampland program~\cite{Vafa:2005ui, Palti:2019pca, vanBeest:2021lhn, Grana:2021zvf} can provide partial answers to these questions by addressing effective field theories with gravity. It advocates the use of the known machinery of general relativity and quantum field theory to conjecture general statements about consistency under the guidance of limited regions of moduli space. 

In this paper, we follow a different route. We focus on the gravity side to explore string-inspired effective actions in settings without supersymmetry, thus using string theory as a guide to constrain their content. In our analysis, we shall also encounter sources without a UV interpretation and a clear fate. They might well become inconsistent once higher order corrections are taken into account, but they could also represent novel types of objects, in the spirit of~\cite{McNamara:2019rup}.

Our main references are Sugimoto's USp(32) model~\cite{Sugimoto:1999tx} with ``brane supersymmetry breaking''~\cite{Antoniadis:1999xk, Angelantonj:1999jh, Aldazabal:1999jr, Angelantonj:1999ms} and the 0'B model~\cite{Sagnotti:1995ga, Sagnotti:1996qj}, which arise as orientifold projections~\cite{Sagnotti:1987tw, Pradisi:1988xd, Horava:1989vt, Horava:1989ga, Bianchi:1990yu, Bianchi:1990tb, Bianchi:1991eu, Sagnotti:1992qw} of the type IIB and type 0B strings~\cite{Seiberg:1986by}. 
These are tachyon-free models in ten dimensions, and their prototype solutions~\cite{Dudas:2000ff} display some of the difficulties ascribable to the lack of supersymmetry. For instance, although perturbative stability holds in some cases~\cite{Basile:2018irz}, non-perturbative effects are unknown and curvature singularities, together with occasional strong coupling regions, cast doubts on their ultimate consistency.

Our contribution will be to examine codimension-one objects in string models not arising from supersymmetry, which are domain walls in a ten-dimensional spacetime interpolating between different vacua.\footnote{A similar analysis can be performed for other branes. The 7-brane case has been recently studied in~\cite{Blumenhagen:2022mqw}. Part of that paper, that appeared while this was in preparation, probes brane constructions similar to ours, but the main focus is different since they are concerned with T-dual versions of what we study here.} Three different cases will be the subject of our investigations, and each will occupy a different section.

In~\cref{sec:uncharged} we start with a toy model where only gravity and the dilaton are turned on. We use that section to fix our notation and to display some of the critical issues we shall encounter later.
In \cref{sec:D8_branes} we focus on (the bosonic part of) massive IIA supergravity and its sources, as indicated by string theory~\cite{Polchinski:1995df}, looking for non-supersymmetric deformations of D8 branes. In the same section we shall also notice peculiar analogies with the effective action of the non-supersymmetric $\text{so}(16)\times\text{so}(16)$ heterotic model in ten dimensions~\cite{AlvarezGaume:1986jb, Dixon:1986iz}. 
In~\cref{sec:Sugimoto_vacua} we turn to our main subject: the two non-supersymmetric orientifold models and their codimension-one vacua. After a brief review of known solutions, we reframe them in the presence of sources, look for generalizations and present the construction in simpler terms.

\section{Uncharged solutions}\label{sec:uncharged}

As explained in the Introduction, we begin by looking for solutions of the Einstein-dilaton equations in 10 dimensions, that follow from the Einstein-frame action
\begin{eqaed}\label{eq:Einstdil}
S=\frac{1}{2\alpha'^4}\int d^{10}x \, \sqrt{g}\left[ R-\frac{1}{2} (\d\phi)^2\right]\mperiod 
\end{eqaed}
We are interested in cases where a nine-dimensional Poincar\'{e} isometry remains. This translates into the codimension-one ansatz
\begin{eqaed}\label{eq:codim1ansatz}
    ds^2&=e^{2A(y)}\eta_{\mu\nu}dx^\mu dx^\nu+e^{2B(y)}dy^2\mcomma \\ \phi&=\phi(y)\mcomma 
\end{eqaed}
that will be our focus throughout this paper. As is clear from \cref{eq:codim1ansatz}, there remains some freedom to redefine $y$, resulting in different choices of $B$ that we shall call ``gauge fixing'' in order to conform with the literature. 

Taking the above ansatz into account, the equations of motion become
\begin{eqaed}\label{eq:einstdilatoneom}
     A'' +A'(9A'-B') &=0 \mcomma \\
     A'' +(A')^2-A'B' +\frac{1}{18}(\phi')^2 &= 0 \mcomma \\
     \phi''+\phi'(9A'-B') &=0\mperiod  
\end{eqaed}
From the first two it is clear that $\phi'=\pm 12 A'$, and one can also set $B=0$ with a suitable choice of the $y$ variable.
The trivial flat case is a solution, as expected, but there is also a non-trivial option: possibly rescaling the $x$ coordinates, one can see that
\begin{eqaed}\label{eq:naiveeq}
     A =\frac{1}{9}\log(y_0\pm 9 y)\mcomma \qquad
     \phi =\phi_0\pm \frac{4}{3}\log(y_0\pm 9y)
\end{eqaed}
solve eqs.~\eqref{eq:einstdilatoneom}.
The double sign inside the logarithm is meant to emphasize that both choices are viable, and, as we shall see, one can combine them to describe extended sources. Note that in $\phi$ there is an independent sign ambiguity.

Taking for instance the positive signs in the arguments, spacetime would end at $y=-y_0/9$ and the curvature scalar would diverge as the inverse of the squared distance from that point. The solution would then extend to arbitrary positive values of $y$ with decreasing scalar curvature. The two choices available for $\phi$ grant two distinct behaviors for the ``string coupling'' $e^\phi$: with the upper one it diverges at infinity, while with the more interesting lower one it diverges at the curvature singularity and vanishes at infinity.

The timelike curvature singularity and the remaining isometries have the flavor of a localized source with $(8+1)$ dimensions. In order to investigate the possible presence and the properties of this 8-brane, one can add a localized contribution to the action and solve the resulting equations. We fix our notation by calling $s(\phi)$ the brane coupling in the Einstein frame:
\begin{eqaed}\label{eq:source_string_frame}
-\int d^9x  \, \sqrt{-\gamma} \, s(\phi) \mperiod
\end{eqaed}
In the string frame\footnote{In this paper our conventions are such that the metric in the string frame is $e^{\frac{1}{2}\phi}$ times that in the Einstein frame, and not $e^{\frac{1}{2}(\phi-\phi_0)}$. The latter would be more appropriate for branes with lower dimensions, where the asymptotic value of the dilaton is a well-defined concept. Clearly, both provide the same results once we turn to the string frame.}, this would become $s(\phi)e^{-\frac{9}{4}\phi}$.

In the following, we shall always work in the Einstein frame and then translate the results into the string-frame in search for a possible string interpretation.

The addition of the localized source in \cref{eq:source_string_frame} to the action makes the first derivatives of the metric and the dilaton not continuous. The equations of motion with a nine-dimensional defect located at $y=0$ are indeed eqs.~\eqref{eq:einstdilatoneom} with a localized source contribution, but remarkably the relation $\phi'=\pm 12 A'$ still holds and leads to $s'(\phi(0))\propto s(\phi(0))$. In principle nothing determines the functional form of $s(\phi)$, but we shall focus on exponential couplings, partly because they make the above condition more natural, but also because perturbative dilaton couplings from string theory usually have this form. The equations of motion are then equivalent to
\begin{eqaed}\label{eq:Aandf}
     A''+9(A')^2=-\frac{1}{16}\delta(y)s(\phi)\mcomma \qquad
     s(\phi)= \tau e^{\mp \frac{3}{4} \phi} \mperiod 
\end{eqaed}

The coupling $s(\phi)$ can have two possible forms, whose counterparts in string frame are
\begin{eqaed}\label{eq:couplings}
e^{-3\phi},\qquad e^{-\frac{3}{2} \phi}\mperiod
\end{eqaed}
These are not perturbative (open or closed) string couplings and therefore one cannot argue for a string origin of this object.
Nevertheless, solving \cref{eq:Aandf} and requiring that $A$ be continuous leads to 
\begin{eqaed}\label{eq:A_uncharged}
   A=\frac{1}{9}\log(y_0 \pm 9|y|)\mcomma
\end{eqaed}
now valid for $y\in\R$, and the jump discontinuity becomes
\begin{eqaed}\label{eq:tensionuncharged}
   \pm \frac{2}{y_0}=-\frac{\tau}{16}e^{\mp \frac{3}{4}\phi_0}\frac{1}{y_0}\mperiod
\end{eqaed}
In this fashion, one is gluing two different vacua, one with $y_0-9y$ and the other with $y_0+9y$. This perspective will become relevant in the following sections. The overall sign choice for $\tau$ in \cref{eq:tensionuncharged} descends from the choices of \cref{eq:A_uncharged} on the two sides of the defect.

An additional issue in \cref{eq:tensionuncharged} is that the tension depends on $\phi_0$. That is also troublesome for a string interpretation, where one would expect no $\phi_0$ dependence.\footnote{Our convention for the Einstein frame in instrumental to fully understand this statement.}
If one wanted to take these defects seriously, they would correspond to some exotic 8-branes, outside the realm of string perturbation theory. They might afford an explanation in terms of cobordism defects in type II theories, along the lines of~\cite{McNamara:2019rup}, but our construction does not interpolate between two constant-dilaton vacua, which makes the interpretation as domain walls at best not transparent.

Let us conclude this section by stressing that the inclusion of a non-vanishing $y_0$ has an additional physical consequence: it makes the solution not divergent at the source. That is reminiscent of what happens for D branes in string theory, which host curvature singularities at the source for $p<8$ (excluding the peculiar $p=3$ case), while the D8 metric is smooth at the brane location, with a curvature singularity at a finite distance from it.

\subsection{D dimensions}

One can also consider the toy model in a general number of dimensions, still with a codimension-one ansatz. 
The equations of motion with a source coupling $s(\phi)$ are a simple generalization of eqs.~\eqref{eq:einstdilatoneom} with localized contributions.
Insisting on an exponential coupling, the solution takes the form
\begin{eqaed}
    s&= \tau \exp{\mp \sqrt{\frac{D-1}{2(D-2)}} \, \phi}  \mcomma \\
    A&=  \frac{1}{D-1}\log\left[y_0\pm(D-1)|y|\right]\mcomma\\
    \phi&=\phi_0\pm\sqrt{\frac{2(D-2)}{D-1}}\log\left[y_0\pm(D-1)|y|\right]\mperiod
\end{eqaed}
Amusingly, the square root is a rational number if\footnote{After playing with prime factors and solving a Pell's equation.}
\begin{itemize}
    \item $D$ is even and
    \begin{eqaed}
        D=\frac{1}{4}\left[\left(3-2\sqrt{2}\right)^{2n}+\left(3+2\sqrt{2}\right)^{2n}+6\right]
    \end{eqaed}
    for $n\in\Z$, so that $D=10,290,\ldots$
    \item $D$ is odd and
    \begin{eqaed}
        D=\frac{1}{8}\left[\left(2+\sqrt{2}\right)\left(3-2\sqrt{2}\right)^n+\left(2-\sqrt{2}\right)\left(3+2\sqrt{2}\right)^n\right]^2+1
    \end{eqaed}
    for $n\in\Z$, so that $D=3,51,\ldots$
\end{itemize}
It is intriguing to see that $D=10$ is singled out somehow from these considerations.

\subsection{Spherically symmetric solution}

As a warm-up exercise for what we shall see in~\cref{sec:Sugimoto_cosmology}, let us now consider the Euclidean version of our problem, in the more general case of a curved nine-dimensional manifold, so that
\begin{eqaed}\label{eq:metricspherically}
    ds^2&=e^{2A(r)}g_{mn}(x)dx^m dx^n+e^{2B(r)}dr^2\mcomma \\
    \phi&=\phi(r) \mperiod
\end{eqaed}

The equations of motion from the action~\eqref{eq:Einstdil} imply that $g_{mn}$ must be the metric on an Einstein manifold. Let us thus define $R_{mn}^{(9)}=8\Lambda_9 g_{mn}$, so that for the sphere $\Lambda_9$ would be the inverse of the squared radius, and let us work in a different gauge, $B=9A$. The dilaton equation remains the same as in~\eqref{eq:einstdilatoneom}, so that
\begin{eqaed}\label{eq:remaining_eom_spherically_symm}
    \phi=\phi_0+\phi_1 r \mcomma \qquad (8 A')^2-64\Lambda_9 e^{16 A}=c_1^2 \mcomma
\end{eqaed}
where $c_1=\frac{2}{3}| \phi_1|$.

In particular, for $\Lambda_9=0$ one thus recovers the results of \cref{sec:uncharged}.
If $\Lambda_9>0$, after adding an appropriate constant to $A$, \cref{eq:remaining_eom_spherically_symm} becomes of the form 
\begin{eqaed}
    (f')^2-1=c_1^2 f^2 \mcomma
\end{eqaed}
where $f=e^{-8A}$, and consequently
\begin{eqaed}
    A=-\frac{1}{8}\log\left[\dfrac{8 \sqrt{\Lambda_9}\sinh\left(c_1 r+r_0\right)}{c_1}\right] \mperiod
\end{eqaed}
When $\Lambda_9<0$ a similar argument leads to 
\begin{eqaed}
    A=-\frac{1}{8}\log\left[\dfrac{8 \sqrt{-\Lambda_9}\cosh\left(c_1 r+r_0 \right)}{c_1}\right] \mperiod
\end{eqaed}
For example, the $\Lambda_9>0$ case with a nine-dimensional sphere interpolates between a flat spacetime and a singular metric of the form $du^2+u^{\frac{2}{9}}d\Omega_9^2$.

\section{Massive IIA, D8 branes and non-supersymmetric analogies}\label{sec:D8_branes}

Let us now turn our attention to a different string-inspired construction. The type IIA string theory has charged codimension-one sources, which are indeed BPS D8 branes and O8 planes acting as domain walls between vacua with different Romans masses~\cite{Romans:1985tz} in the bulk.

The effective actions of interest involve gravity, dilaton and the R-R nine-form of type IIA.
Source terms describing the low energy effective contribution of branes and orientifolds are nine-dimensional integrals with a coupling to the R-R field and a tension term.

The equations of motion with sources are to be supplemented with Bianchi identities. Their content is the counterpart of R-R tadpole cancellations in 2d CFTs and they enforce a global R-R charge cancellation for compact solutions. In fact, non-zero Romans mass arises precisely in the absence of local R-R charge cancellation. It is a piecewise constant function between charged codimension-one sources, whose jump discontinuities are dictated by
\begin{eqaed}\label{eq:Romans_mass_normalization}
    dm_0=-2\dsum_i q_i\delta(y-y_i)\mcomma
\end{eqaed}
where we included a factor of $\sqrt{\alpha'}$ in the definition of the R-R charges $q_i$ with respect to the standard notation, so as to avoid $\alpha'$ factors in the following equations. 

After integrating out the R-R field, the Einstein-frame action of interest, with the addition of a D8 brane-like source at $y=0$ and with our codimension-one ansatz, becomes
\begin{eqaed}\label{eq:Einstein_frame_IIA}
    S=\frac{1}{2(\alpha')^4}\dint d^{10}x \,  \sqrt{g}\left[R-\frac{1}{2}(\d\phi)^2-2\frac{q^2}{16}e^{\frac{5}{2}\phi}- \tau e^{\frac{5}{4}\phi} \frac{\delta(y)}{\sqrt{g_{yy}}}\right] \mperiod
\end{eqaed}
The reason for taking out a factor of $2$ will become clear in \cref{sec:heterotic_analogy}. In what follows, we shall always isolate the factors $\frac{q}{4}$ for the same purpose.

\subsection{Bulk analysis}

Our first step is now to find all possible vacua of \cref{eq:Einstein_frame_IIA} without sources with the ansatz in \cref{eq:codim1ansatz}, which may be even generalized by considering a nine-dimensional Ricci-flat metric instead of the flat one. We already know one example, the D8 brane, which in the Einstein frame takes the form
\begin{eqaed}\label{eq:D8_brane}
    ds^2 & = \left(1-h_8 y\right)^{\frac{1}{8}}dx_{(9)}^2+\left(1-h_8 y\right)^{\frac{9}{8}}dy^2 \mcomma \\
    e^\phi & = \left(1-h_8 y\right)^{-\frac{5}{4}}\mperiod
\end{eqaed}

In our codimension-one setting, the equations of motion can be presented as
\begin{eqaed}\label{eq:IIA_eom_on_ansatz}
    A''+(9A'-B')A'+\frac{1}{4}\frac{q^2}{16}e^{2B+\frac{5}{2}\phi} & = 0 \mcomma \\
    144 (A')^2-(\phi')^2+4\frac{q^2}{16}e^{2B+\frac{5}{2}\phi} & = 0 \mcomma \\
    \phi''+(9A'-B')\phi'-5\frac{q^2}{16}e^{2B+\frac{5}{2}\phi} & = 0\mperiod
\end{eqaed}
It is certainly possible to use the $B=9A$ gauge, so as to recover the D8 brane solution in~\eqref{eq:D8_brane} and to try to explore whether generalizations exist. However, we shall take a different route, also inspired by the way we have written eqs.~\eqref{eq:IIA_eom_on_ansatz}.

The gravity and dilaton terms in the action of massive IIA, after integrating out the R-R field, are the same as in one of the tachyon-free non-supersymmetric string models in ten dimensions, specifically the $\text{so}(16)\times\text{so}(16)$ heterotic model. This simple comment is actually convenient for our setup since the techniques of~\cite{Dudas:2000ff} are then available. For that reason, we choose our coordinate $y$ such that $B=-\frac{5}{4}\phi$ and define
\begin{eqaed}\label{eq:good_redefinition}
    f(y)=\log\left(\sqrt{1+\frac{16}{q^2}  (6A')^2}+\frac{4}{{q}}(6A')\right)\mcomma
\end{eqaed}
in terms of which $A'$ and $\phi'$ become\footnote{We shall remain agnostic about the sign of $q$. In fact, both cases of $q>0$ and $q<0$ in our notation are meaningful, as they imply that every solution with $y_0+y$ has a partner solution with $y_0-y$. Had we worked with $|q|$, an additional $\pm$ sign ambiguity would make the notation more cumbersome.}
\begin{eqaed}\label{eq:A_prime_and_phi_prime}
    A' =\frac{1}{6}\frac{{q}}{4}\sinh f\mcomma \qquad
    \phi' =\pm 2\frac{{q}}{4}\cosh f\mperiod
\end{eqaed}
The equations of motion reduce to
\begin{eqaed}\label{eq:f_function_10D}
    f'+\frac{3}{2}\frac{{q}}{4}\cosh{f}\pm\frac{5}{2}\frac{{q}}{4} \sinh{f} = 0\mperiod
\end{eqaed}

The simplest $f$ that satisfies \cref{eq:f_function_10D} is actually $f=\mp\log 2$, where the signs are in one-to-one correspondence with those in \cref{eq:f_function_10D} and \cref{eq:A_prime_and_phi_prime}.
The two solutions in the bulk are thus
\begin{eqaed}\label{eq:massive_IIA_solutions_1}
    ds^2 & = e^{\mp\frac{1}{4}\frac{{q}}{4}y} dx_{(9)}^2+e^{-\frac{5}{2}\phi_0}e^{\mp\frac{25}{4}\frac{{q}}{4}y} dy^2 \mcomma \\
    e^\phi & = e^{\phi_0}e^{\pm\frac{5}{2}\frac{{q}}{4}y}\mperiod
\end{eqaed}
One can recast these in the gauge $B=9A$, resorting to the following reparametrization, up to additive constants:
\begin{eqaed}
    z=\frac{1}{2}\frac{4}{{q}} e^{-\frac{5}{4}\phi_0}e^{\mp 2 \frac{{q}}{4}y}\mperiod
\end{eqaed}
Both signs, related by a coordinate transformation, yield the D8 brane bulk spacetime
\begin{eqaed}\label{eq:D8_solution_1}
    ds^2 & = \left[2\frac{{q}}{4}e^{\frac{5}{4}\phi_0} z \right]^{\frac{1}{8}} dx_{(9)}^2 + \left[2\frac{{q}}{4}e^{\frac{5}{4}\phi_0}z \right]^{\frac{9}{8}}dz^2 \mcomma \\
    e^\phi & = e^{\phi_0} \left[2\frac{{q}}{4}e^{\frac{5}{4}\phi_0} z\right]^{-\frac{5}{4}}\mperiod
\end{eqaed}

For the second type of non-constant $f$ satisfying  \cref{eq:f_function_10D}, we refer to~\cite{Dudas:2000ff} for the derivation and simply quote the result, using $\frac{|q|}{4}$ instead of $\sqrt{\beta_{\mbox{\tiny E}}}$ in that paper. The basic relation is
\begin{eqaed}\label{eq:f_solution_generic}
    e^f = \pm 2^{\mp 1}\frac{e^{\frac{{q}}{4}y} + \varepsilon e^{-\frac{{q}}{4}y}}{e^{\frac{{q}}{4}y} - \varepsilon e^{-\frac{{q}}{4}y}} \mcomma
\end{eqaed}
where $\varepsilon$, without loss of generality, can be only $\pm1$. When one computes $A'$ and $\phi'$ using \cref{eq:A_prime_and_phi_prime}, the sign ambiguity of~\eqref{eq:f_solution_generic} results in a sign flip for $\varepsilon$, so that one can pass from one choice to the other by simply sending $\varepsilon\to -\varepsilon$. We decide to keep the upper sign in what follows, and then, if $\varepsilon=1$
\begin{eqaed}\label{eq:massive_IIA_solutions_2}
    ds^2 & = \left(\sinh{\frac{{q}}{4}y}\right)^{\frac{1}{12}} \left(\cosh{\frac{{q}}{4}y}\right)^{-\frac{1}{3}}dx_{(9)}^2+ e^{-\frac{5}{2}\phi_0}\left(\sinh{\frac{{q}}{4}y}\right)^{-\frac{5}{4}}\left(\cosh{\frac{{q}}{4}y}\right)^{-5} dy^2\mcomma\\
    e^{\phi} & = e^{\phi_0} \left(\sinh{\frac{{q}}{4}y}\right)^{\frac{1}{2}}\left(\cosh{\frac{{q}}{4}y}\right)^{2}\mperiod
\end{eqaed}
If $\varepsilon=-1$, the resulting solution is \cref{eq:massive_IIA_solutions_2}, after interchanging $\cosh$ with $\sinh$.
In the gauge $B=9A$, letting
\begin{eqaed}
    z=-\frac{4}{{q}}e^{-\frac{5}{4}\phi_0}\log\tanh\left(\frac{q}{4} y \right)\mcomma
\end{eqaed}
the two solutions are
\begin{eqaed}\label{eq:D8_solution_3}
    ds^2 & = \exp{\mp\frac{5}{24}\frac{{q}}{4} e^{\frac{5}{4}\phi_0}z}\left[2\sinh\left(\frac{{q}}{4} e^{\frac{5}{4}\phi_0}z\right)\right]^{\frac{1}{8}} dx_{(9)}^2 + \\ & + \exp{\mp\frac{15}{8}\frac{{q}}{4} e^{\frac{5}{4}\phi_0}z}\left[2\sinh\left(\frac{{q}}{4} e^{\frac{5}{4}\phi_0}z\right)\right]^{\frac{9}{8}} dz^2 \mcomma \\
    e^{\phi} & = e^{\phi_0} \exp{\pm\frac{3}{4}\frac{{q}}{4} e^{\frac{5}{4}\phi_0}z}\left[2\sinh\left(\frac{{q}}{4} e^{\frac{5}{4}\phi_0}z\right)\right]^{-\frac{5}{4}}\mperiod
\end{eqaed}
In this form, these were already displayed in~\cite{Mourad:2021qwf}.

\subsection{A comment on a heterotic case}\label{sec:heterotic_analogy}

As we already stressed, the bulk equations of motion for gravity and dilaton in massive IIA, after integrating out the R-R 9-form, coincide with those of the non-supersymmetric $\text{so}(16)\times\text{so}(16)$ heterotic model in ten dimensions~\cite{AlvarezGaume:1986jb, Dixon:1986iz}. Using the notations of~\cite{Dudas:2000ff}, our IIA solutions can thus be mapped to the heterotic ones letting $\frac{|q|}{4}\to\sqrt{\beta_{\mbox{\tiny E}}}$. Since we kept $\frac{q}{4}$ explicitly in the expressions for metrics and dilatons, it is straightforward to compare the two cases.

We can actually add something that was not noticed in the original work for the $\text{so}(16)\times\text{so}(16)$ heterotic vacua, namely the existence of a solution with $f'=0$, \cref{eq:D8_solution_1}.\footnote{This exists in any dilaton-gravity model with an exponential potential whose exponent is greater than the critical one.} It corresponds to the D8 brane in type IIA, therefore describes a defect whose coupling to the dilaton is $e^{-\phi}$ in string frame. By adding an integration constant and reintroducing the sign ambiguity, one can recover the known properties of the D8 solution. While the mapping exists, it is unclear how this consideration can play a role in the non-supersymmetric heterotic case. No D branes exist in that model, so that the $e^{-\phi}$ coupling is not singled out by string arguments, and so far this result is merely an artifact of the lowest-order terms in the effective actions.

\subsection{Sources in the deformed solutions}

Returning to type IIA, let us investigate the possible addition of 8-branes to the three new solutions that we have found (for \cref{eq:D8_solution_1} we already know the answer). To this end, we add in eqs.~\eqref{eq:IIA_eom_on_ansatz} the source term from \cref{eq:Einstein_frame_IIA}, still in the convenient gauge $B=-\frac{5}{4}\phi$.

Taking into account \cref{eq:massive_IIA_solutions_2}, one must demand that $A$ and $
\phi$ be continuous, while enforcing, for the derivatives, the jump discontinuities
\begin{eqaed}\label{eq:IIA_sources_only_tadpole_gauge}
    \Delta A'  = - \frac{1}{16}\tau \mcomma \qquad
    \Delta \phi'  = \frac{5}{4}\tau \mperiod
\end{eqaed}
From \cref{eq:IIA_sources_only_tadpole_gauge} one would need $\Delta\phi'=-20\Delta A'$, but, as the reader may notice from the exponents of \cref{eq:massive_IIA_solutions_2}, this condition turns out to be equivalent to demanding $\left(\cosh y_0\right)^2 = \left(\sinh y_0\right)^2$, which rules out this source.

There remains the possibility of gluing different types of solutions. We shall not glue the D8 to other solutions since it has a physical meaning by itself. Rather, we focus on the two non-trivial vacua in \cref{eq:massive_IIA_solutions_2}, where the result will be physically more interesting. We take for $y>0$
\begin{eqaed}
    A & = A_1 + \frac{1}{24}\log\cosh{\left(y_1+\frac{q}{4}y\right)}-\frac{1}{6}\log\sinh{\left(y_1+\frac{q}{4}y\right)}\mcomma \\
    \phi & = \phi_1+\frac{1}{2}\log\cosh{\left(y_1+\frac{q}{4}y\right)}+2\log\sinh{\left(y_1+\frac{q}{4}y\right)}\mcomma
\end{eqaed}
and for $y<0$
\begin{eqaed}
    A & = A_0 + \frac{1}{24}\log\sinh{\left(y_0-\frac{q}{4}y\right)}-\frac{1}{6}\log\cosh{\left(y_0-\frac{q}{4}y\right)}\mcomma \\
    \phi & = \phi_0+\frac{1}{2}\log\sinh{\left(y_0-\frac{q}{4}y\right)}+2\log\cosh{\left(y_0-\frac{q}{4}y\right)}\mperiod
\end{eqaed}
Continuity fixes $A_1$ and $\phi_1$ in terms of the other parameters, while one can set $A_0=0$ with a global redefinition of $x_\mu$. The condition $\Delta\phi'=-20\Delta A'$ becomes simply $y_1=y_0$, while the other matching in \cref{eq:IIA_sources_only_tadpole_gauge} is
\begin{eqaed}\label{eq:D8_deformed_tension}
    \tau = \frac{\cosh{(2y_0)}}{\sinh{(2y_0)}} q\mperiod
\end{eqaed}
Two sign options are actually available in \cref{eq:D8_deformed_tension}, depending on how one glues the two solutions of \cref{eq:massive_IIA_solutions_2}, but in any case this condition tells us that $|\tau|>|q|$. 

Wrapping up, in this section we have found two possible sources consistent with the equations of motion of massive IIA. One is the D8 brane, while the other is a different object, still coupled with $e^{-\phi}$ in the string frame. We have once more no string interpretation, but at least the object behaves as if it were some kind of non-supersymmetric deformation of a D8 brane.
This deformed source has two fundamental properties that one should expect for sources in the absence of supersymmetry, a tension greater than charge signaling a potential instability for decay into a supersymmetric D8, and the gravitational backreaction that forbids multiple static sources.

\section{Vacua for the ten-dimensional orientifolds}\label{sec:Sugimoto_vacua}

In this section we turn to the main topic of this paper, which is the search for possible codimension-one variants of the Dudas-Mourad vacua for orientifold models. These are solutions to the equations of motion of the two ten-dimensional tachyon-free orientifold strings without supersymmetry: the USp(32) model~\cite{Sugimoto:1999tx} and the 0'B model~\cite{Sagnotti:1995ga, Sagnotti:1996qj}. They have the same low-energy effective actions, insofar as only gravity and the dilaton are taken into account, and for definiteness we shall refer to the former, which affords an interpretation in terms of more familiar (anti) BPS branes and orientifolds, mirroring the usual type I construction.
After a brief review of their known codimension-one vacua, we shall be concerned with 8-branes in these models, along the lines of what we did in \cref{sec:uncharged}.

The USp(32) model arises as a different orientifold projection from the type IIB string. In the spacetime picture, it includes an orientifold O9$^+$ and 32 $\overbar{\mbox{D9}}$ branes. The total R-R charge vanishes, so that the model is anomaly free,\footnote{However, the possible presence of global anomalies is still an open problem. We thank I. Basile and A. Debray for discussions on the topic.} but there is a leftover NS-NS contribution, which results in a scalar potential for the dilaton proportional to $e^{-\phi}$ in the string frame, hinting at the open string origin of ``brane supersymmetry breaking''~\cite{Antoniadis:1999xk, Angelantonj:1999jh, Aldazabal:1999jr, Angelantonj:1999ms}.
We shall not review the construction here, but refer to~\cite{Angelantonj:2002ct, Mourad:2017rrl} for more details.

Adding a scalar potential has a dramatic consequence: the classical background cannot be a ten-dimensional Minkowski spacetime. However, the string-generated potential for the dilaton can conspire with fluxes to yield $AdS$ solutions with a stabilized $\phi$~\cite{Mourad:2016xbk, Basile:2018irz, Raucci:2022bjw}. 

Without fluxes turned on, one must give up something to find a solution, as can be seen from purely dimensional reasons: the scalar potential $\propto \int T e^{-\phi}$ carries an extra factor $1/\alpha'$ with respect to the closed string contributions. Therefore, $T$ must enter the solution together with some of the spacetime coordinates, which signals the necessary breaking of the full ten-dimensional Poincar\'{e} symmetry. This result was originally found by Dudas and Mourad in~\cite{Dudas:2000ff}, and we now briefly review it.

\subsection{The Dudas-Mourad orientifold vacua}

In the Einstein frame, the effective action for the orientifold models of~\cite{Sugimoto:1999tx, Sagnotti:1995ga, Sagnotti:1996qj}, considering only metric and dilaton contributions, takes the form\footnote{We use the definition of $\alpha_{\mbox{\tiny E}}$ from~\cite{Dudas:2000ff}. In the USp(32) model it is the sum of tensions from (anti) BPS branes and orientifold, multiplied by $\alpha'^4$, and similarly with the 0'B model, where branes and orientifolds are not BPS.} 
\begin{eqaed}\label{eq:Dudas_Mourad_action}
    S=\frac{1}{2\alpha'^4}\int d^{10}x \,  \sqrt{-g}\left[R-\frac{1}{2}(\d\phi)^2-2\alpha_{\mbox{\tiny E}} e^{\frac{3}{2}\phi}\right]\mperiod
\end{eqaed}
With the usual ansatz in \cref{eq:codim1ansatz}, choosing the gauge $B=-\frac{3}{4}\phi$, the equations of motion become
\begin{eqaed}\label{eq:DM_equations_of_motion}
     A''+9(A')^2+\frac{3}{4}A'\phi'+\frac{1}{4}\alpha_{\mbox{\tiny E}}  &= 0\mcomma \\
     72(A')^2-\frac{1}{2}(\phi')^2+2\alpha_{\mbox{\tiny E}} &= 0\mcomma \\
     \phi''+\phi'\left(9A'+\frac{3}{4}\phi'\right)-3\alpha_{\mbox{\tiny E}} &= 0\mperiod
\end{eqaed}
Letting
\begin{eqaed}\label{eq:good_redefinition_in_10D}
    f(y)=\log\left(\sqrt{1+\frac{36 (A')^2}{{\alpha_{\mbox{\tiny E}}}}}+\frac{6A'}{\sqrt{\alpha_{\mbox{\tiny E}}}}\right)\mcomma
\end{eqaed}
one of the equations becomes clearly redundant and one is left with 
\begin{eqaed}\label{eq:A_prime_and_phi_prime_10D}
    A' =\frac{\sqrt{\alpha_{\mbox{\tiny E}}}}{6}\sinh f\mcomma \qquad
    \phi' =\pm 2\sqrt{\alpha_{\mbox{\tiny E}}}\cosh f\mperiod
\end{eqaed}
A single non-trivial condition for $f$ follows, with a sign ambiguity inherited from \cref{eq:A_prime_and_phi_prime_10D}:
\begin{eqaed}\label{eq:f_function_10D_orientifold}
    2f'+3\sqrt{\alpha_{\mbox{\tiny E}}}\cosh f \pm 3\sqrt{\alpha_{\mbox{\tiny E}}}\sinh f =0\mcomma
\end{eqaed}
solved by
\begin{eqaed}\label{eq:f_equation_10D_orientifold}
    f=\mp\log\left(y_0\pm\frac{3}{2}\sqrt{\alpha_{\mbox{\tiny E}}} \, y\right)\mperiod 
\end{eqaed}
One is free to choose either the upper or lower signs, but this ambiguity disappears after one integration to obtain $A$ and $\phi$, from which one can write metric and dilaton as (for $y>0$ and with $y_0=0$)
\begin{eqaed}\label{eq:metric_and_dilaton_orientifold}
    ds^2 & = (\sqrt{\alpha_{\mbox{\tiny E}}} \, y)^{\frac{1}{9}}e^{-\frac{1}{8}\alpha_{\mbox{\scalebox{.4}{E}}} y^2}dx_{(9)}^2+ e^{-\frac{3}{2}\phi_0}(\sqrt{\alpha_{\mbox{\tiny E}}} \, y)^{-1}e^{-\frac{9}{8}\alpha_{\mbox{\scalebox{.4}{E}}} y^2}dy^2 \mcomma \\
    e^{\phi}& = e^{\phi_0} (\sqrt{\alpha_{\mbox{\tiny E}}} \, y)^{\frac{2}{3}} e^{\frac{3}{4}\alpha_{\mbox{\scalebox{.4}{E}}} y^2}\mperiod
\end{eqaed}
This is the solution originally found in~\cite{Dudas:2000ff}, in which two timelike curvature singularities exist at $y=0$ and $y\to\infty$, and the proper length in the internal $y$ direction is finite. The string coupling $e^\phi$ vanishes at $y=0$ and diverges at infinity.

From a physical perspective, since the proper length is finite, one would like to interpret \cref{eq:metric_and_dilaton_orientifold} as a metric on an interval, including the two singular endpoints as end-of-the-world defects. The singularities are timelike, and it is natural to regard them as two codimension-one branes, backreacting on the ten-dimensional geometry.
This interpretation is supported by various arguments involving the pinch-off singularity at finite distance~\cite{Antonelli:2019nar, Buratti:2021yia, Buratti:2021fiv}, but no conclusive answer from string theory is known.

In what follows, we shall address the problem by explicitly adding a source term to the action.

\subsection{Gluing with branes}

The sign choice in \cref{eq:f_equation_10D_orientifold} allows one to glue two vacua with opposite signs and different parameters $y_0$ and $y_1$. To this end, one must deform the action, including a localized source with a generic coupling $s(\phi)$, so that
\begin{eqaed}\label{eq:sugimoto_action_with_source}
    S\propto\int d^{10}x \,  \sqrt{-g}\left[R-\frac{1}{2}(\d\phi)^2-2 {\alpha_{\mbox{\tiny E}}} e^{\frac{3}{2} \phi}\right]-\int d^{9}x \,  \sqrt{-\gamma}  \,  s(\phi)\mcomma
\end{eqaed}
up to the common $2\alpha'^4$ factor.

The equations of motion with the source at $y=0$ are those in eqs.~\eqref{eq:DM_equations_of_motion} with an additional localized contribution, which introduces jump discontinuities in the first derivatives of $A$ and $\phi$:
\begin{eqaed}\label{eq:sugimoto_jumps}
    \Delta A'=-\frac{1}{16}e^{-\frac{3}{4}\phi(0)}s(\phi(0))\mcomma \qquad \Delta\phi'=e^{-\frac{3}{4}\phi(0)}s'(\phi(0))\mperiod
\end{eqaed}

We should understand how to glue the two vacua and whether one can regard the resulting geometry as a domain wall separating them. One is free to choose one of \cref{eq:f_equation_10D_orientifold} for $y<0$ and the other for $y>0$. With one choice, our bulk solution is, in the region $\frac{3}{2}\sqrt{\alpha_{\mbox{\tiny E}}} \, y+y_0>0$,
\begin{eqaed}\label{eq:Sugimoto_bulk_0}
    A & = A_0 -\frac{1}{16}\left(\sqrt{\alpha_{\mbox{\tiny E}}} \, y+\frac{2}{3}y_0\right)^2 +\frac{1}{18}\log\left(\sqrt{\alpha_{\mbox{\tiny E}}} \, y+\frac{2}{3}y_0\right) \mcomma \\
    \phi & = \phi_0+\frac{3}{4}\left(\sqrt{\alpha_{\mbox{\tiny E}}} \, y+\frac{2}{3}y_0\right)^2+\frac{2}{3}\log\left(\sqrt{\alpha_{\mbox{\tiny E}}} \, y+\frac{2}{3}y_0\right) \mperiod
\end{eqaed}
The other choice applies to the region $y_1-\frac{3}{2}\sqrt{\alpha_{\mbox{\tiny E}}} \, y>0$ and can be written as
\begin{eqaed}\label{eq:Sugimoto_bulk_1}
    A & = A_1 -\frac{1}{16}\left(\frac{2}{3}y_1-\sqrt{\alpha_{\mbox{\tiny E}}} \, y\right)^2 +\frac{1}{18}\log\left(\frac{2}{3}y_1-\sqrt{\alpha_{\mbox{\tiny E}}} \, y\right) \mcomma \\
    \phi & = \phi_1+\frac{3}{4}\left(\frac{2}{3}y_1-\sqrt{\alpha_{\mbox{\tiny E}}} \, y\right)^2+\frac{2}{3}\log\left(\frac{2}{3}y_1-\sqrt{\alpha_{\mbox{\tiny E}}} \, y\right) \mperiod
\end{eqaed}
One can remove an additive constant in $A$ by performing an overall redefinition of the spacetime coordinates $x_\mu$ on the two sides of the source. 

We now make two additional assumptions in order to explicitly identify a source term. The first one is $A_0=A_1=0$, so that the continuity of $A$ translates into the relation
\begin{eqaed}\label{eq:y0y1_relation}
    y_0^2-2\log y_0=y_1^2-2\log y_1\mcomma
\end{eqaed}
between $y_0$ and $y_1$, with two possible solutions. One solution is $y_1=y_0$, and guided by the expectation that $y_{0,1}$ should have a physical meaning related to the source we select this as the second assumption, but the reader should be aware of this subtlety. Then, the continuity of $\phi$ at $y=0$ fixes $\phi_1=\phi_0$. 
Jump discontinuities are still present in the first derivatives, but there will be differences with respect to \cref{sec:uncharged}. 

Let us work out the case of an exponential source $s(\phi)=\tau e^{\beta\phi}$, taking for instance \cref{eq:Sugimoto_bulk_0} for $y>0$ and \cref{eq:Sugimoto_bulk_1} for $y<0$ (we need $y_0>0$). Matching the discontinuities gives the conditions
\begin{eqaed}\label{eq:delta_matching}
    -\frac{1}{6}\left(y_0-\frac{1}{y_0}\right)\sqrt{\alpha_{\mbox{\tiny E}}}& = -\frac{\tau}{16} e^{\beta\phi(0)}e^{-\frac{3}{4}\phi_0}e^{-\frac{1}{4}y_0^2}\sqrt{\frac{3}{2y_0}}\mcomma \\
    2\left(y_0+\frac{1}{y_0}\right)\sqrt{\alpha_{\mbox{\tiny E}}}&=\beta \tau e^{\beta\phi(0)}e^{-\frac{3}{4}\phi_0}e^{-\frac{1}{4}y_0^2}\sqrt{\frac{3}{2y_0}}\mperiod
\end{eqaed}
Hence, the complete coupling to the dilaton is
\begin{eqaed}\label{eq:Sugimoto_source_coupling}
    s(\phi)=\tau \exp{\frac{3}{4}\frac{y_0+\frac{1}{y_0}}{y_0-\frac{1}{y_0}}\phi}\mcomma
\end{eqaed}
with an explicit dependence on the parameter $y_0$, and the tension is given by 
\begin{eqaed}
    \tau = \frac{8}{3}\sqrt{\alpha_{\mbox{\tiny E}}}\left(y_0-\frac{1}{y_0}\right)\exp{\frac{1}{y_0^2 -1}\left[-\frac{3}{2}\phi_0-\frac{1}{2}y_0^2-\log\left(\frac{2}{3} y_0\right)\right]}\mperiod
\end{eqaed}
The coupling is not completely fixed by the equations of motion, and the ambiguities arise since $A'$ and $\phi'$ are not proportional. The dilaton potential rules out $\phi'\propto A'$ in the equations of motion, making the matching non-trivial.

The reader should note how the tension depends on the dilaton zero-mode and on $\alpha'$.
Recall that there is a factor of $2\alpha'^4$ hidden in $\tau$ and that, in our conventions, $\alpha_{\mbox{\tiny E}}\sim (\alpha')^{-1}$ without powers of $\phi_0$ (there would be an $e^{-\phi_0}$ in the other Einstein frame convention). Hence, in string frame 
\begin{eqaed}
    s(\phi)\sim (\alpha')^{-\frac{9}{2}} e^{-\frac{3}{2}\phi_0}\exp{\frac{3}{4}\left(\frac{y_0+\frac{1}{y_0}}{y_0-\frac{1}{y_0}} - 3 \right)(\phi-\phi_0)}\mcomma
\end{eqaed}
which should be compared with $(\alpha')^{-\frac{9}{2}} e^{-\phi_0}e^{-(\phi-\phi_0)}$, the expression that would apply to BPS D branes.
The unusual dependence of the tension of $\phi_0$ leads us to believe that $\phi_0$ may be a measure of how many sources are backreacting on the geometry.

The coupling can become proportional to $e^{-\phi}$ in the string frame if $y_0=2$, but obtaining no $\phi_0$ dependence in the tension is impossible.
Alternatively, one could allow $y_0$ to depend on $\phi_0$, so as to cancel the $\phi_0$ dependence in $\tau$, but then one would be left with a complicated exponential coupling to the dilaton from \cref{eq:Sugimoto_source_coupling}.

In view of these difficulties, one could be tempted to relax some assumptions that we made, in particular not demanding that $A_1=A_0$ (therefore $y_1\neq y_0$), while insisting that the tension be independent of $\phi_0$. Using \cref{eq:sugimoto_jumps} with an exponential coupling, a natural choice is
\begin{eqaed}
    s(\phi)=\tau e^{\frac{3}{4}\phi}\mperiod
\end{eqaed}
However, the jump discontinuities would become inconsistent since they require that $y_0^{-1}+y_1^{-1}=0$, while $y_{0,1}>0$ for the solution to exist. Alternatively, the relevant coupling for D branes, that is $s(\phi)=\tau e^{\frac{5}{4}\phi}$, can be attained by requiring $y_0 \, y_1=4$, while allowing a complicated functional form for $y_0$ in terms of $\phi_0$.

To summarize the content of this section, the sources that we find in the Dudas-Mourad orientifold vacua have peculiar properties. They rule out a perturbative string origin, although near $y=0$ neither high curvature nor strong coupling are generically present. The most disturbing feature is probably the dependence of the tension on $\phi_0$, which calls for a better understanding of the role of $\phi_0$ in these vacua.

\subsection{Remarks on curved manifolds}\label{sec:Sugimoto_cosmology}

Let us now turn to the Euclidean equations of motion for the orientifold models, where the ansatz becomes \cref{eq:metricspherically}.
Since $g_{mn}$ must be the metric on an Einstein manifold, we define $R_{mn}=\Lambda_9 g_{mn}$.

The approach that we are about to follow can be summarized by saying that we try to ``compensate the tadpole with internal curvature''.
This can be done by restricting our ansatz so that\footnote{This is an ansatz, and not a gauge choice. In fact, we are not fixing $B$ for the moment.}
\begin{eqaed}\label{eq:cancel_tadpole_with_curvature}
    A = -\frac{3}{4}\phi\mcomma
\end{eqaed}
which simplifies the equations, turning them into
\begin{eqaed}\label{eq:sad_equation}
    \Lambda_9+2\alpha_{\mbox{\tiny E}} = 0 \mcomma \qquad (\phi')^2=-\frac{\alpha_{\mbox{\tiny E}}}{2} e^{\frac{3}{2}\phi+2B}\mperiod 
\end{eqaed}
The latter is inconsistent with the sign of the tadpole potential, and therefore a solution of this type does not exist.

Interestingly, however, these steps become relevant in a time-dependent setting, where
\begin{eqaed}\label{eq:time_depentend_metric}
    ds^2 & = - e^{2B(t)}dt^2 + e^{-\frac{3}{2}\phi(t)}g_{mn}(x)dx^m dx^n\mcomma \\
    \phi  &= \phi(t)\mperiod
\end{eqaed}
Some signs change in the equations of motion with respect to the Euclidean case, because of the time signature.
The two cases differ only in the contributions of $\alpha_{\mbox{\tiny E}}$ and $\Lambda_9$, and the solution now exists.
Summarizing, we have an internal hyperbolic space with $\Lambda_9=-2\alpha_{\mbox{\tiny E}}$ and
\begin{eqaed}\label{eq:cosmological_solution}
    A   = -\frac{3}{4}\phi\mcomma \qquad
    (\phi')^2 =\frac{\alpha_{\mbox{\tiny E}}}{2} e^{\frac{3}{2}\phi+2B} \mperiod
\end{eqaed}
Note that, even before gauge-fixing, the metric can be expressed in terms of $\phi$ alone as
\begin{eqaed}\label{eq:not_gauge_fixed_cosmological}
    ds^2=e^{-\frac{3}{2}\phi}\left[-\frac{2}{\alpha_{\mbox{\tiny E}}}(\phi')^2 dt^2+g_{mn}dx^m dx^n\right]\mperiod
\end{eqaed}
For instance, in the $B=0$ gauge 
\begin{eqaed}\label{eq:cosmo_gauge_2}
    ds^2 & = -dt^2+\left(t_0\pm\frac{3}{4}\sqrt{\frac{\alpha_{\mbox{\tiny E}}}{2}} \, t\right)^2 g_{mn}dx^m dx^n\mcomma \\
    e^\phi & = \left(t_0\pm\frac{3}{4}\sqrt{\frac{\alpha_{\mbox{\tiny E}}}{2}} \, t\right)^{-\frac{4}{3}}\mperiod
\end{eqaed}

\subsection{A tale of frames}\label{sec:frames}

Working in the Einstein frame is convenient both from a practical perspective, because the gravity equations are simpler and more familiar, and from a physical perspective, because the Einstein frame metric contains only gravitational degrees of freedom while the dilaton has a canonical kinetic term. On the other hand, the equations of motion in string theory arise naturally in string frame. Nevertheless, is the Einstein frame the best option from a computational standpoint?
In this section, we show that for the ten-dimensional orientifold models another frame may be useful.

One can indeed simplify the exponential potential in \cref{eq:Dudas_Mourad_action} by a change of frame. If $g_{MN}^{\mbox{\tiny E}}$ is the Einstein-frame metric, letting
\begin{eqaed}\label{eq:new_frame}
    g_{MN}=e^{\frac{3}{2}\phi}g_{MN}^{\mbox{\tiny E}}
\end{eqaed}
the action takes the form
\begin{eqaed}\label{eq:action_new_frame}
    S=\int d^{10}x  \, \sqrt{g} \, e^{-6\phi}\left[R+40(\d\phi)^2-2\alpha_{\mbox{\tiny E}}\right]\mcomma
\end{eqaed}
and the equations of motion, after some simplifications, become
\begin{eqaed}\label{eq:equations_new_frame}
    R_{MN}+6\nabla_M \d_N\phi+4\d_M\phi \d_N\phi + 2\alpha_{\mbox{\tiny E}} g_{MN} & = 0\mcomma \\
    \Box\phi-6(\d\phi)^2-3\alpha_{\mbox{\tiny E}} & = 0\mperiod
\end{eqaed}
There are no dilaton exponentials in the tadpole terms of these equations, but the metric equation is more complicated than in the Einstein frame. Note, however, that the Ricci scalar is simply 
\begin{eqaed}\label{eq:Ricci_scalar_new_frame}
    R=-40(\d\phi)^2-38\alpha_{\mbox{\tiny E}}\mperiod
\end{eqaed}

We want to see how the new cosmological solution of \cref{sec:Sugimoto_cosmology} and the Dudas-Mourad vacua emerge in this framework. 

The first of eqs.~\eqref{eq:equations_new_frame} simplifies considerably if $R_{mn}=-2\alpha_{\mbox{\tiny E}} g_{mn}$, and therefore let us concentrate on a product metric with a nine-dimensional $g_{mn}$ and no warp factors, without specifying the signature of the remaining direction. 
The Ricci scalar is then $R=-18 \alpha_{\mbox{\tiny E}}$ and \cref{eq:Ricci_scalar_new_frame} reduces to
\begin{eqaed}\label{eq:new_frame_cosmo_equations}
    (\d\phi)^2=-\frac{1}{2}\alpha_{\mbox{\tiny E}}\mperiod
\end{eqaed}
This calls for a time-dependent $\phi$, and we are thus led to
\begin{eqaed}\label{eq:new_frame_time_dependent_ansatz}
    ds^2  &= -dt^2+g_{mn}dx^m dx^n\mcomma\\
    \phi  &= \phi(t)\mcomma
\end{eqaed}
only constrained by \cref{eq:new_frame_cosmo_equations}. This is the cosmological solution of \cref{sec:Sugimoto_cosmology}, here recovered in a simpler fashion.

The Dudas-Mourad solution is not as simple, but the final form of the metric is still simpler than that in the Einstein frame. 
Let us choose our coordinates so that $B=0$ in \cref{eq:codim1ansatz}, and define the following two combinations
\begin{eqaed}
    X=\frac{3}{2}A-\phi \mcomma \qquad  Y=2A-\frac{5}{3}\phi \mcomma
\end{eqaed}
which reduce eqs.~\eqref{eq:equations_new_frame} to
\begin{eqaed}\label{eq:new_frame_DM_equations_redef}
    X''+6(X')^2  =0\mcomma \qquad
%    Y''+\frac{\alpha_{\mbox{\tiny E}}}{2}  =0\mcomma \qquad
    X'Y'+\frac{\alpha_{\mbox{\tiny E}}}{12}  =0\mperiod
\end{eqaed}
The final expression for the bulk metric and dilaton is
\begin{eqaed}\label{eq:new_frame_DM}
    ds^2  &= e^{\frac{3}{2}\phi_0} (\sqrt{\alpha_{\mbox{\tiny E}}} \, y)^{\frac{10}{9}}e^{ \alpha_{\mbox{\scalebox{.4}{E}}} y^2}\eta_{\mu\nu}dx^\mu dx^\nu + dy^2 \mcomma \\
    e^\phi  &= e^{\phi_0} (\sqrt{\alpha_{\mbox{\tiny E}}} \, y)^{\frac{2}{3}} e^{\frac{3}{4} \alpha_{\mbox{\scalebox{.4}{E}}} y^2} \mperiod
\end{eqaed}
This is indeed the Dudas-Mourad solution in this frame, as can be seen by comparing with \cref{eq:metric_and_dilaton_orientifold}.
We hope to have convinced the reader with these simple examples that our different frame choice simplifies computations in the presence of a tadpole potential. 

A natural question would be to ask whether this is still true beyond codimension-one cases. 
For instance, as a generalization, one could consider metrics that are fibered over an interval. This has been explored recently~\cite{Mourad:2021roa} and, as the reader can verify, many solutions without form fluxes in that work take a simpler form in this frame.
Unfortunately, \cref{eq:new_frame} does not simplify terms involving R-R fluxes. In the language of~\cite{Mourad:2021qwf, Mourad:2021roa} the contributions from R-R $(p+2)$-form field strengths are accompanied by a dilaton exponential
\begin{eqaed}\label{eq:R-R_dilaton_coupling}
    \exp{\left(-2\beta_p +\frac{3}{2}(p+1)\right)\phi}\mcomma 
\end{eqaed}
and no relevant simplifications arise since $\beta_p=\frac{p-3}{4}$, both for the USp(32) model and for the type 0'B model. Hence, a $(p+2)$-forms will carry an $e^{(p+3)\phi}$ factor, and solving the equations in the presence of the tadpole potential is not any simpler than solving them in the potential-free case.

\section{Conclusions}

In this paper we have studied codimension-one vacua and defects interpolating between them, drawing some inspiration from string theory models. These are subjects that are attracting increasing attention in the string literature, in particular due to the Swampland program. 
Our gravitational analysis may be useful in complementing the works of~\cite{Buratti:2021yia, Buratti:2021fiv, Angius:2022aeq, Blumenhagen:2022mqw} about end-of-the-world branes, and consequently those of~\cite{Basile:2021mkd, Basile:2022zee}, where the aim was to verify Swampland conjectures in the models of interest here. One could also consider generalizations of these settings with internal tori, along the lines of~\cite{Maxfield:2014wea}.

Our investigation of codimension-one sources has not led to conclusive answers, as was to be expected, since a full string theory analysis is impossible with present technology. It would also be interesting to extend the analysis to T-dual versions of the non-supersymmetric ten-dimensional strings, even with no local R-R tadpole cancellation, starting from~\cite{Blumenhagen:2000dc, Dudas:2002dg}. The work of~\cite{Blumenhagen:2022mqw} is a first step in this direction.

In \cref{sec:D8_branes} we have also found sources that could represent non-supersymmetric deformation of D8 branes. Although these objects lack, at present, a proper string theory interpretation, in codimension one the equations of motion and the supersymmetry conditions have very similar content, which allows some potentially useful steps. In fact, we have extended the work of~\cite{Dudas:2000ff} for the non-supersymmetric $\text{so}(16)\times\text{so}(16)$ model with a previously unnoticed solution. 

While our solutions have a meaning in field theory, it is unclear whether they should be recognized as physically relevant, in particular because of their unexpected source couplings. The idea we want to convey is that our knowledge of the supersymmetric vacua can provide in principle some insight into models without supersymmetry. In our study, that meant using similarities between portions of the $\text{so}(16)\times\text{so}(16)$ effective action and that of massive IIA. This particular map is lost once one considers more terms, for instance the NS-NS 2-form field, but we are currently investigating how to extend the strategy to other models, in particular those of \cref{sec:Sugimoto_vacua}.

Although it is possible to explore the perturbative stability of the new vacua that we have found, there is apparently no natural way to address non-perturbative stability. 
However, some simple cases have proved tractable, as in the case of brane nucleation instabilities in vacua with R-R fluxes~\cite{Antonelli:2019nar}.
Similarly, sources of instability like bubbles of nothing~\cite{Witten:1981gj} could exist for the flux-less solutions~\cite{GarciaEtxebarria:2020xsr}, and the Dudas-Mourad vacuum, while perturbatively stable~\cite{Basile:2018irz}, might suffer from non-perturbative instability.\footnote{Witten's Kaluza-Klein instability will play a role after an additional internal $S^1$ compactification, because any nine-dimensional Ricci-flat metric solves the equations of motion. But here we are referring to the stability of the codimension-one solution as such.} 
Further tools to explore solutions in models without supersymmetry may provide new strategies to investigate stability, along the lines of~\cite{Giri:2021eob}.

\acknowledgments

We are grateful to A. Sagnotti for numerous comments on the manuscript. We thank I. Basile for stimulating discussions, and R. Blumenhagen, N. Cribiori, C. Kneissl and A. Makridou for explanations on their related work. 
This work was supported in part by Scuola Normale Superiore, by INFN (IS GSS-Pi) and by the MIUR-PRIN contract 2017CC72MK\_003.

\bibliographystyle{JHEP}
\bibliography{codim1}

\end{document}